\documentclass[
    ,final            
  ]
  {aipproc}

\layoutstyle{8x11single}

\begin{document}

\title{Tracking and vertexing at ATLAS}


\author{Pamela Ferrari \\
   On behalf of the ATLAS collaboration}{
  address={CERN, PH division, 1211 Geneve 23 Switzerland}
}

\begin{abstract}

Several algorithms 
for tracking and for primary and secondary vertex reconstruction have
been developed by the ATLAS collaboration 
following different approaches. This has allowed a thorough
cross-check 
of the performances of the algorithms and of the reconstruction software. The results of
the most recent studies on this topic are discussed and compared.

\end{abstract}

\maketitle


\section{Introduction}

ATLAS is one of the four experiments at the Large
Hadron Collider (LHC) that will produce $pp$ collisions
 at a centre-of-mass energy of 14 TeV.
There will be a bunch crossing every 25 ns, which imposes
fast response from the detectors as well as the need to store data
from each collision in on-detector pipelines until a decision of the 
first level of the trigger is taken. 
Furthermore there will be a very high density of charged tracks: at
the nominal luminosity of 10$^{34}$cm$^{-2}$sec$^{-1}$ for each bunch
crossing there will be about 200 charged tracks and about 15 vertex candidates.
The reconstruction of tracks and of primary and secondary vertices at LHC will
be a challenging task.
The Inner Detector (ID)
is  the tracking detector of ATLAS and consists of three
sub-detectors, whose performances as simulated by the software are
extensively described in ~\cite{IDTDRI,IDTDRII}.
The outermost detector  is the Transition Radiation Tracker (TRT) consisting of
several layers of 4 mm straws in the barrel region (arranged in 3 layers 
of modules) and 14 Transition Radiation Tracker wheels in the endcap, 
providing  about 30 hits per track  and  a resolution in the $R\phi$
plane of 170 $\mu m$.
The SemiConductor Tracker (SCT) is located in the region between 
25 cm to 50 cm in radius  and consists of 
4 layers of stereo silicon strips detectors in the barrel and 9 disks 
per side in the endcaps, achieving a resolution of about 17 $\mu m$ and 580 $\mu m$
in the $R\phi$ and $Rz$ plane, respectively.
The innermost is the pixel detector 
~\cite{pixel}, consisting of  
3 layers of silicon pixel detectors (at radius of 5.05, 8.85 and 12.25 cm) in
the barrel and 3 disks in each of the endcaps, providing a track 
resolution of about 11 $\mu m$ in R$\phi$ and 60 $\mu m$ in $Rz$. 
To reconstruct the tracks, it is essential to take into account 
the effect of multiple scattering and of the energy loss in the material,
which requires a precise knowledge of material in the detector.
More complicated tracking algorithms are needed to consider the track
resolution degradation at the edges of the ID due to the non-uniformity 
of the 2T magnetic field provided by the ATLAS solenoid in the ID region.

\section{Tracking and vertexing at trigger level }

The ATLAS trigger is subdivided into three different trigger
selection layers that  reduce the 1 GHz interaction rate to 200 Hz.
\begin{itemize}
\item The first level trigger (LVL1) is a hardware trigger  with a 2.5 ms
  latency that brings the rate to 75kHz.
It uses reduced granularity data of the calorimeter and the muon
  detectors and identifies  geometrical regions of interest (RoI)
  in the detector. 
\item The second level of the trigger (LVL2) uses  the ID
 information since it  processes in parallel
 the full information of  all sub-detectors in the RoIs defined by LVL1.
 It has a latency of 10 ms bringing the rate below 2 kHz.
Tracking and vertexing can be performed at LVL2.
\item Finally the Event Filter (EF) can use algorithms similar to the 
offline software having a 2s latency time.

\end{itemize}
The performance of the tracking and vertexing algorithms implemented at ATLAS 
at LVL2 is described below.

\subsection{Tracking at LVL2}

The tracking algorithm forms track seeds by fitting with a straight 
line pairs of space points in the pixel innermost layer (B layer) and in the 
second logical layer (in a given RoI).
The tracks are extrapolated back to the beam line and an
Impact Parameter (IP) is obtained for each of them.
The track is retained if the IP in the transverse plane with respect to
the beam axis is small.
The $z$ coordinate of the primary vertex is then obtained as the maximum of the
histogram filled with the $z$ intersection of the seeds with the beam line.
A third space point is extracted in modules situated in positions
where the hits may lay based on the track extrapolation.
After having removed the ambiguities due to overlapping space points 
in triplets using the extrapolation quality, the triplets are fitted
and identified  with tracks. The efficiency for tracks in jets is
80-90\% depending   on the  luminosity and the  event topology 
and 95\% for single electrons.

\subsection{$b$-tagging  at LVL2}

The $b$-jet selection is performed by using the transverse 
impact parameter $d_0$ significance, 
defined as $S = d_0/s(d_0)$, where $s(d_0$) is the
error on  $d_0$ and its dependence on $p_T$ is obtained from the
simulation.

A secondary vertex algorithm similar to offline but faster is used.
The $b$-jet estimator uses a likelihood ratio given by the 
product of the ratios of the probability densities
for each track to come from a $b$-jet or a light-jet,
$$W=\Pi_i \frac{f_b(S_i)}{f_u(S_i)}.$$
where $i$ runs over all reconstructed tracks in the jet.

The final discriminative variable 
is defined as  X = W/(1+W). The processing time for $b$-jets is less than 2 ms.
The rejection for light quarks is 25\% and 15\% for a $b$-jet efficiency of 
50\% and 60\% respectively.

\section{offline tracking algorithms}

Several offline tracking  algorithms have been developed at ATLAS:
xKalman and iPatRec and only very recently the NewTracking algorithm.
\begin{itemize}

\item The xKalman first searches for tracks in the TRT using fast
  histogramming of straw hits, then it extrapolates back to SCT and
  pixels and fits the tracks using a  Kalman filter that
  associates clusters to tracks assuming that the noise and all
  material effects and measurements are Gaussian.
  Finally, the improved tracks are extrapolated back into the TRT in a 
  narrow region around the extrapolated trajectory, retaining all hits
  in that region. 
\item the iPatRec algorithm forms track-candidates in SCT and pixel detectors
using space-point combinatorials subject to criteria on maximum curvature and crude
vertex region projectivity. A global $\chi^2$ fitter is used to fit
tracks and associate clusters.
Only good tracks are retained for extrapolation in TRT, where TRT hits
are added.
To limit the contamination from high occupancy, tight cuts are applied
on the straw residuals.
\item The NewTracking algorithm is mainly a reorganization of
  the tracking code. At present it is largely based on xKalman, but in
  the future it will use also tools from iPatRec with the aim of obtaining
  optimised performances. Furthermore it uses a 
  better detector geometry description.
\end{itemize}   
Table~\ref{tab:trackalg} shows the results of a performance comparison
done using $t\bar{t}$ events.
As can be seen xKalman and iPatRec have comparable performances,
slightly better than the newly developed NewTracking, whose
performances are going to improve and are already better than the
others at the time of writing.

\begin{table}
\begin{tabular}{cccc}
\hline
  & \tablehead{1}{c}{b}{xKalman}
  & \tablehead{1}{c}{b}{iPatRec}
  & \tablehead{1}{c}{b}{NewTracking}   \\
\hline
Multiplicity (p>1 GeV) & 16.69 & 17.06 & 16.88\\
Barrel Track eff/fake rate &	99\%/0.6\% &	99\%/0.7\% &
  96\%/2.5\% \\
Transition eff/fake rate &	98\%/0.6\%  &	98\%/0.5\% &
  96\%/3.6\% \\
Forward eff/fake rate &	98\%/0.3\% &	99\%/1.3\% &	95\%/2.7\% \\
\hline
\end{tabular}
\caption{ Performance of the ATLAS offline tracking algorithms
 obtained using $t\bar{t}$ events. 
The result of the newly developed NewTracking are very preliminary and 
already superseeded at the time of writing.}
\label{tab:trackalg}
\end{table}

\begin{figure}
  \includegraphics[height=.3\textheight]{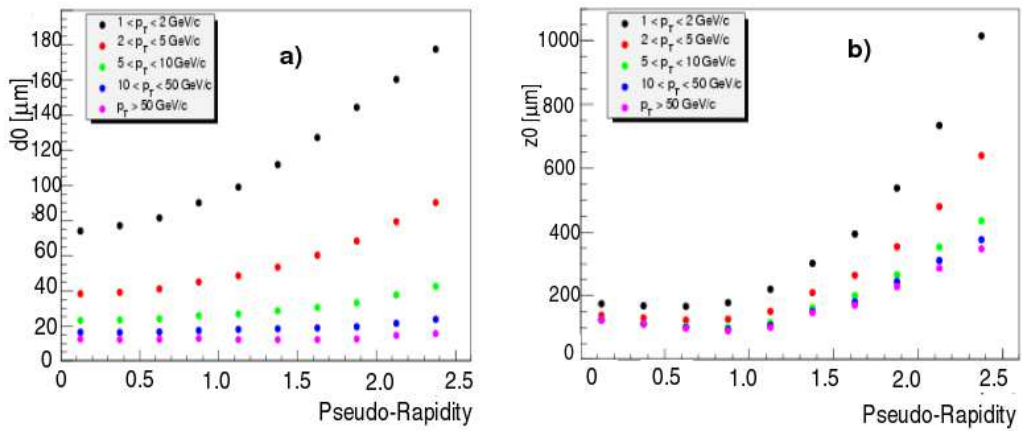}
  \caption{Figure a) and b) show the transverse and longitudinal 
impact parameter resolution for a sample of WH events with $m_H$= 400 $GeV/c^2$,
where the W is decaying to $\mu\nu_{\mu}$ and the H to $uu$.}
  \label{fig:fig1}
\end{figure}

\begin{figure}
  \includegraphics[height=.4\textheight]{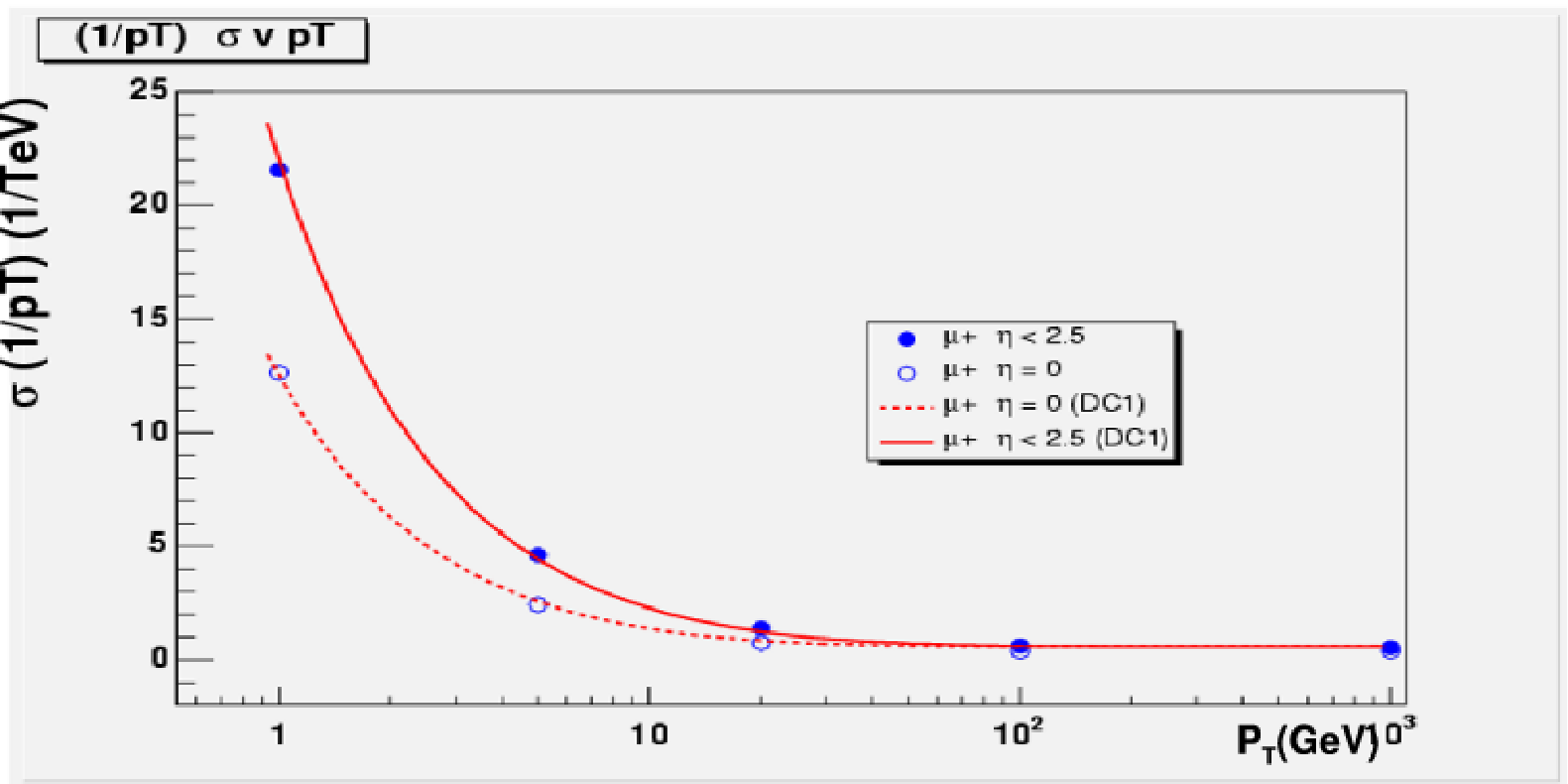}
  \caption{The momentum resolution for single muons versus $p_T$
averaged over  all $|\eta|$ (full circles) and for $\eta$=0 (open
circles). The full lines show results obtained with an older 
simulation considering only 2 pixel layers instead than 3.
    }
  \label{fig:pttrack}
\end{figure}

Figures~\ref{fig:fig1} a) and b) show the resolution on the impact 
parameter in the transverse and longitudinal planes, respectively.
A sample of WH events with m$_H$= 400 $GeV/c^2$ has been used for the
study, where the W decays to $\mu\nu_{\mu}$ and H to $uu$. 
Figure~\ref{fig:pttrack} shows the momentum resolution versus $p_T$ for single muons
averaged over  all $|\eta|$ and for $\eta$=0.


\section{Primary vertex reconstruction}

Given the large multiplicity of tracks at each bunch crossing 
(several hundreds as discussed before) the vertex reconstruction must be fast
and robust.
The input consists of  the 3 dimentional  trajectories  and error
matrices of the tracks. Quality requirements on tracks are applied:
typically p$_T$ > 1 GeV/c, |d$_0$| < 0.25 mm, |z$_0$| < 150 mm and
$\chi^2$ per track <0.5.

The approximate primary vertex position in $z$ is found using a  
sliding window of 0.7 cm that is moved along the whole interaction
region. The window with largest number of tracks, weighted with $p_T$
is chosen. The <$z$> position of the vertex is given by the mean
of all the tracks in that window.
Tracks belonging to the primary vertex are taken away and the procedure is 
iterated to get other (pile-up) vertices.

All tracks at $\pm$ 5 mm in $z$ and $\pm$ 1 mm in  the transverse plane
are accepted as coming from primary vertex.
At this point the vertex fitting is performed using a Billoir
method. In the fitting procedure the outliers are removed, that is if the 
$\chi^2$ obtained when adding a track is too high, the track is
rejected and the fit is recalculated, 

There are two different implementations of this method which are
basically using the same strategy: VxPrimary and
VKalVrt.

A third primary vertex fitter has been developed,
the Adaptive Vertex Fitter, that solves the problem 
of outlier tracks that spoil the fit, not by 
discarding them, but by down-weighting them.
It minimises instead than residuals, the sum 
of squared residuals weighted with the $\chi^2$.
 Table~\ref{tab:primver} shows a comparison of the performances of the
primary vertex finder algorithms, which give very
similar results by following slightly different approaches. 

\begin{table}
\begin{tabular}{ccc}
\hline
  & \tablehead{1}{c}{b}{x($\mu m$)}
  & \tablehead{1}{c}{b}{z($\mu m$)}   \\
\hline
VxPrimary & 12.6 $\pm$ 0.1 & 50.0 $\pm$ 0.5 \\
AVF  & 11.07 $\pm$ 0.09 &	46.76 $\pm$ 0.05 \\
VKalVrt & 11.07$\pm$0.09 &	45.43 $\pm$ 0.05 \\
\hline
\end{tabular}
\caption{ Primary vertex resolution in $x$ and $z$ for the ATLAS
primary vertex finders. A sample of simulated WH events with
m$_H$=120 GeV/$c^2$ and $W\rightarrow\mu\nu_{\mu}$, $H\rightarrow
b\bar{b}$ has been used for the study.}
\label{tab:primver}
\end{table}

\section{ATLAS $b$-tagging}

Basic ingredients for the $b$-tagging are the long lifetime of the
$b$-hadron jets, the high multiplicity $b$-jets and the impact parameters
of the tracks.
Additionally the presence of a secondary vertex can increase the 
rejection power of the $b$ selection.
The $b$-tagging algorithms developed by the ATLAS collaboration are
based on such properties of the $b$-jet~\cite{btag}.
In addition $b$-tagging methods based on the tagging of a 
soft lepton  (electron or muon) resulting from the decay of a 
$b$-flavoured hadron are being developed.

The standard $b$-tagging identification is based on the 2-dimentional
impact parameter of a track in the transverse plane.
The method adopted by ATLAS uses the normalised significance $S=d_0/s(d_0)$
 for each track and
compares it to predefined calibration probability density functions 
for the $b$ and light quark hypothesis to obtain the probabilities 
$b(S)$ and $u(S)$. 
The following discriminating variable (IP2D) for each jet is built by summing 
over all the tracks in a jet
$$ W_{jet}=\sum_{i=1}^{N_{tr}} ln \frac{b(S_i)}{u(S_i)}.$$
The discriminative variable consists of logarithms of
the density functions ratio, allowing to combine two variables easily.
The $b$-tagging performance can be improved by adding the longitudinal 
and the transverse significance (IP3D):
$$ W= \frac{P_b(S_{d_0},S_{z_0})}{P_u(S_{d_0},S_{z_0})}.$$

\begin{figure}
  \includegraphics[height=.3\textheight,width=\textwidth]{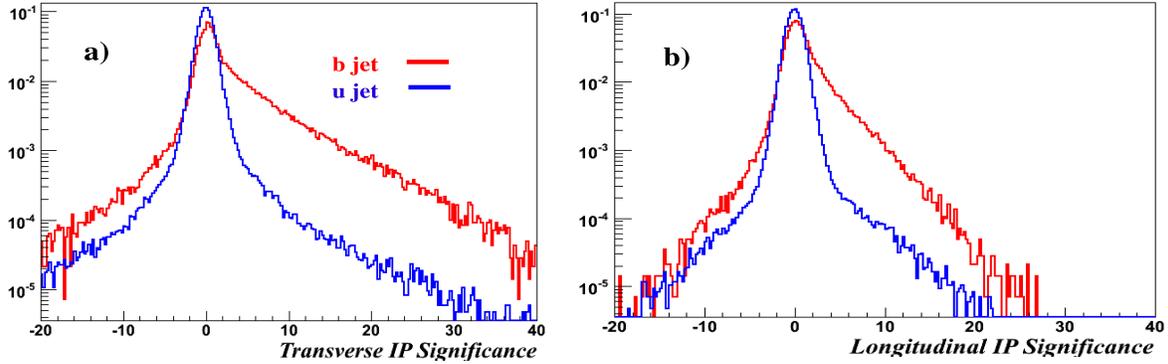}
  \caption{Figure a) and b) show the distribution of the 
 transverse and longitudinal impact parameter significance.  }
  \label{fig:IPbtag}
\end{figure}

Figures~\ref{fig:IPbtag} a) and b) show the distribution of the
transverse and longitudinal  impact parameter significance for $b$ 
and light quarks, respectively.

In addition the presence of a secondary vertex is searched for.
Only tracks passing the following quality cuts are used:
$p_T >$ 1 GeV/c, $|\eta|< 2.5$,  $|d_0|< 1$  mm,   $|z_0|< 1.5$ mm, 
number of pixel hits in the B layer greater than 0 and 
number of pixel hits larger
than 1, while more than 6 SCT hits should be detected.
The selected tracks are used to search for good 2 track vertices in
the jet. At this point  vertices that are due to 
decays of Kaons or Lamdba baryons or to interactions with the 
beam pipe material, the pixel layers or to photon
conversions  (V0's) are removed. 
A  common (inclusive) vertex is formed for the  remaining tracks.

\begin{figure}
  \includegraphics[width=0.8\textwidth]{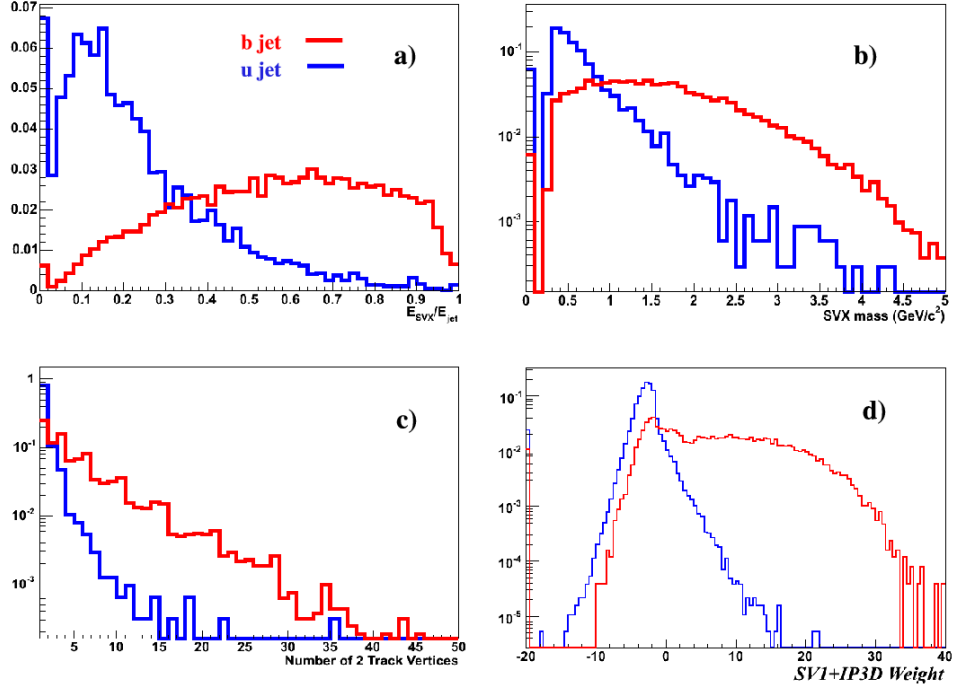}
  \caption{Figure 
a), b), c) show the distribution of the secondary vertex
variables for $b$ and light jets. Figure d) shows the weight obtained 
by combining SV1 and IP3D.}
  \label{fig:secvxt}
\end{figure}

Once the secondary vertex is reconstructed  some additional discriminating
variables can be considered: 
\begin{enumerate}
\item the number of good two track vertices in the jet,
\item invariant mass of all particles coming from the secondary vertex mass,
\item the ratio of the energy of all particles coming from  the secondary vertex and the total energy of the jet.
\end{enumerate} 
The variables have been chosen to be independent from tracks impact 
parameters in order to have a real gain on $b$-tagging performance.
The secondary vertex discriminating variable  
obtained from the probability density functions that parametrize the 
distributions of these  variables can be combined with the track impact
parameter based $b$-tagging procedure.
Figure~\ref{fig:secvxt} a), b and c) show the distribution of the secondary vertex
variables for $b$ and light jets.

A jet is labeled as a $b$-jet if there is a $b$-quark within a cone of radius 
0.3 around the jet axis, the efficiency for $b$-jets ($\epsilon_b$) is defined as the
ratio of the number of jets and  the number of jets labeled as 
$b$ with $p_T > 15 $ GeV/c and $|\eta|<2.5$.
The light jet rejection is defined as $R_u=1/\epsilon_u$.

Overlapping jets cause mislabeling and furthermore the jet isolation is very
dependent on the type of physics process taken into account.
A purification of the jets is done  to factorize those effects
 from pure $b$-tagging issues: light jets are not taken into account if
 there is a $b$/$c$/quark/hadron within a cone of radius 0.8  around the
 jet axis.
Table~\ref{tab:b-tag} shows the rejection for light jets for a $b$-tag
 efficiency of 50\% and 60\% for the IP2D and IP3D impact parameter
 and for the combined IP3D and SV1 discriminating variables.
Figure~\ref{fig:secvxt} shows the combined SV1+IP3D weight.
The results are obtained using a sample of
 WH$\rightarrow  \mu\nu_{\mu}u\bar{u}$ events with m$_H $~= 120 GeV/c$^2$.
Tracks have been reconstructed with the xKalman algorithm and two
 different primary 
vertex finders have been compared: the adaptive vertex fitter and
 VkalVrt.
In the last two rows of the table, results are shown  when rejecting bad tracks 
recognised as coming from V0's or interactions with the beam pipe or
 detector material.
In this study the jets were reconstructed around the primary quark
 directions with a cone size of $\Delta R$= 0.4. 

\begin{table}
\begin{tabular}{lcccccc}
\hline
  & \multicolumn{2}{c} {IP2D}
  & \multicolumn{2}{c}{IP3D}
  & \multicolumn{2}{c}{IP3D+SV1}  \\
\hline
efficiency &	50\%	& 60\% & 50\% &	60\% & 50\%	& 60\% \\
Rej. VKalVrt & 135$\pm$9 &	55 $\pm$ 2 &	214 $\pm$ 18 &
  75 $\pm$ 4	 & 609 $\pm$	86 &	157 $\pm$ 11 \\
Rej. AVF &	130 $\pm$ 9 &	52 $\pm$ 2 & 205 $\pm$ 17 &	73
  $\pm$ 4 & 	612 $\pm$ 87 $\pm$ &	147 $\pm$10 \\
Rej. no VO's + VkalVrt &	206$\pm$17 &	69 $\pm$ 3 &
  339 $\pm$35 &	101 $\pm$6 &	815 $\pm$ 134 &	192 $\pm$ 15 \\
Rej. no V0's + AVF &
	199 $\pm$ 16 & 	66 $\pm$ 3 &	327 $\pm$ 34 &	98 $\pm$6 &
  794 $\pm$129	 & 164 $\pm$ 12 \\
\hline
\end{tabular}
\caption{b-tagging rejection for an efficiency of 50\% and 60\%
for the IP2D, IP3D and SV1 taggers, using the AVF and VkalVrt primary vertex
fitters.  
A sample of WH$\rightarrow  \mu\nu_{\mu}u\bar{u}$ events with m$_H$~= 120
GeV/c$^2$
events has been used for the study.
The last two rows show the results when rejecting tracks coming from
V0's.}
\label{tab:b-tag}
\end{table}

An evaluation of the $b$-tagging performance using 190K of  $t\bar{t}$ events has
shown similar results: the rejection for light quarks obtained using
SV1+IP3D for a 50\% efficiency for $b$-quarks is $858\pm42.9$ while for
a 60\% efficiency is $259\pm7.8$.

\section{Conclusions}

There has been a lot of work and improvements on the tracking and 
vertexing algorithms in the ATLAS collaboration  in the past few years.
Different approaches have been followed in parallel to develop
the  tracking and vertexing algorithms, giving comparable results.
Cosmic events have successfully been reconstructed with the SCT and TRT
barrel on the surface in the spring of this year.



\bibliographystyle{aipproc}   


\end{document}